\documentclass[journal]{IEEEtran}
\usepackage{cite}
\usepackage{amsmath,amssymb,amsfonts}
\usepackage{subcaption}
\usepackage{textcomp}
\usepackage{hyperref}
\usepackage{nicefrac}
\usepackage{soul}
\usepackage{pifont}
\usepackage{xcolor}
\usepackage{cuted}
\usepackage[pdftex]{graphicx}
\usepackage{algorithm}
\usepackage{algpseudocode}
\usepackage{ifthen}
\usepackage{graphicx}
\usepackage{diagbox}

\algnewcommand\algorithmicinput{\textbf{input:}}
\algnewcommand\algorithmicoutput{\textbf{output:}}
\algnewcommand\algorithmicbl{\textbf{\# of blocks:}}
\algnewcommand\algorithmicth{\textbf{\# of threads / block:}}
\algnewcommand\Input{\item[\algorithmicinput]}
\algnewcommand\Output{\item[\algorithmicoutput]}
\algnewcommand\Block{\item[\algorithmicbl]}
\algnewcommand\Thread{\item[\algorithmicth]}
\algrenewcommand\algorithmicindent{1.0em}

\newcommand{\llr}{llr}

\newcommand{\BO}{\alpha_{out}} 

\newcommand{\BI}{\alpha_{in}} 
\newcommand{\BM}{\delta}
\newcommand{\PM}{\sigma}

\newcommand{\SP}{\pi}
\newcommand{\K}{k} 
\newcommand{\B}{\beta} 
\newcommand{\N}{n} 
\newcommand{\F}{f} 
\newcommand{\Ovr}{v} 

\DeclareMathOperator*{\argmax}{arg~max}

\begin{document}
\title{High-Throughput and Memory-Efficient Parallel Viterbi Decoder for Convolutional Codes on GPU}
%
%
%

\author{Alireza~Mohammadidoost~
and~Matin~Hashemi
\thanks{A. Mohammadidoost and M. Hashemi are with the Department of Electrical Engineering, Sharif University of Technology, Tehran, Iran. E-mails: mohammadidoost@ee.sharif.edu, matin@sharif.edu (corresponding author). }}

%

\maketitle

\begin{abstract}
This paper describes a parallel implementation of Viterbi decoding algorithm. Viterbi decoder is widely used in many state-of-the-art wireless systems. The proposed solution optimizes both throughput and memory usage by applying optimizations such as unified kernel implementation and parallel traceback. Experimental evaluations show that the proposed solution achieves higher throughput compared to previous GPU-accelerated solutions. 
\end{abstract}


%
\IEEEpeerreviewmaketitle

\section{Introduction}
\label{sec:intro}
%
%
%
%

\IEEEPARstart{C}{hannel} coding is a technique that is widely employed in transmission of data over an unreliable or noisy communication channel. The transmitter encodes the original message by adding redundancy. This enables the receiver to recover the original data by decoding the received noisy data. 
Convolutional coding is a channel coding method which has been widely used in industrial protocols, for instance, in DVB-T, DVB-S, GPRS, GSM, LTE, 3G, CDMA, WiFi and WiMAX. 
Different decoding algorithms exist for convolutional codes, among which the Viterbi decoding algorithm is the optimal and the most widely-used method.

The Viterbi decoder operates in either hard-decision mode or soft-decision mode. In the hard-decision mode, every bit in the input of the decoder is represented by either a zero or one. In the soft-decision mode, however, every input bit is a log likelihood ratio (LLR) that is formed based on the probability that the received bit is zero or one. 
In this mode, the Viterbi decoder takes advantage of the additional information in order to decrease the bit error rate (BER) by about $2.3$~dB. A lower BER means a better recovery of the original signal. This comes at the cost of higher computational requirement, which in turn, lowers the overall decoding throughput.

Many FPGA-based methods have been proposed for acceleration of the Viterbi decoding algorithm. While such methods achieve very high throughput, they do not provide the flexibility required for software defined radio (SDR) and cognitive radio (CR) applications.

This paper proposes a novel parallel algorithm for implementation of the Viterbi decoding algorithm in the soft-decision mode on GPU hardware. 
Flexibility is a key factor in software defined radio (SDR) and GPU provides a platform for flexible software-based implementations at high throughput.

The proposed solution is mainly focused on optimizing the memory requirement. This aspect has not been explored in previous works but has a very large impact in boosting the throughput. The proposed algorithm reaches a very high decoding rate compared to previous solutions. 

\section{Preliminaries}
\label{sec:prelim}

This section presents a brief overview of convolutional encoding, Viterbi decoding algorithm, and the concept of soft-decision inputs. In addition, a brief overview of CUDA API for parallel programming on GPU hardware is presented.


\begin{figure}[tp]
	\centering
	\includegraphics[width=1.0\columnwidth]{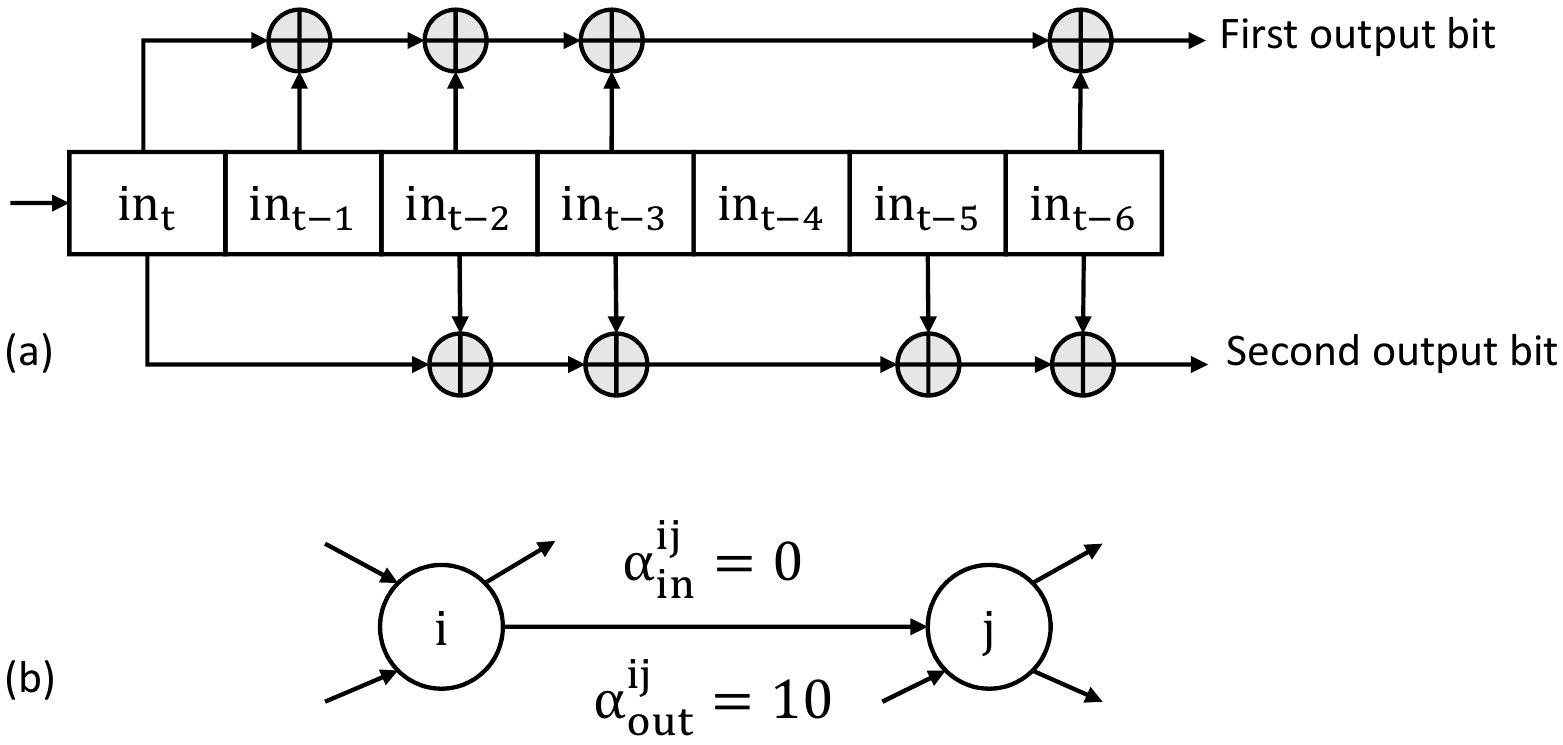} 
	\caption{a) Convolutional encoder $(\B, 1, \K)$ with $\K=7$, $\B=2$, and generator polynomials $1111001$ and $1011011$. b) Branch $ij$ from state $i$ to $j$ in the encoder FSM.}
	\label{fig:encoder}
\end{figure}

\subsection{Convolutional Encoder} 

Fig.~\ref{fig:encoder}(a) shows an example \textbf{convolutional encoder}. The encoder receives a series of $\N$  bits and produces a series of encoded bits which will be transmitted over the communication channel. 
$\B \geq 2$ encoded bits are generated for every input bit. \textbf{Code rate} is defined as the inverse of $\B$. 
At time (stage) $t$, each one of the $\B$ output bits is computed based on the current input bit, i.e., $in_t$, and the previous $\K-1$ input bits as 
\begin{equation}
(g_{\K-1} . in_{t}) \oplus \cdots \oplus (g_0 . in_{t-\K+1}) 
\label{eq:encoder}
\end{equation}
where, $\K$ is called the \textbf{constraint length}, $\oplus$ is the $xor$ operator, and $g$ is a $\K$-bit value called the \textbf{generator polynomial}. There are $\B$ generator polynomials, one for every output bit. 
In the example shown in Fig.~\ref{fig:encoder}(a), $\K=7$ and $\B=2$. The two generator polynomials are $1111001$ and $1011011$, whose octal representations are $171$ and $133$, respectively.

The encoder can be viewed as a finite state machine (FSM) with $2^{\K-1}$ \textbf{states}. The previous $\K-1$ input bits, i.e., $(in_{t-1}, in_{t-2}, \ldots in_{t-\K+1})$, form the current state. 
Assume the FSM is in state $i \in [0,2^{\K-1}-1]$. Depending on $in_t$, i.e., the current input bit which is either zero or one, the FSM takes a \textbf{branch} $ij$ from state $i$ to state $j$. 
Hence, a series of input bits causes the FSM to take a series of branches, which is called a \textbf{path}.

The encoder FSM can be formed solely based on $\K$, $\B$ and the generator polynomials. Every branch $ij$ in the FSM is associated with a single-bit \textbf{branch input} $\BI^{ij}$ and a $\B$-bit \textbf{branch output} $\BO^{ij}$. See Fig.~\ref{fig:encoder}(b). The output of the encoder is $\BO^{ij}$ provided that the FSM goes from state $i$ to $j$. This transition happens if the input bit is equal to $\BI^{ij}$.

\subsection{Viterbi Decoding Algorithm} 

The encoded bits are transmitted over the channel. When they arrive at the receiver, some of the bits are corrupted. The received bits are decoded in order to recover the original data. 
The Viterbi decoding algorithm is the optimal and the most widely-used method for decoding convolutional codes \cite{Viterbi}. 

Given the received series of bits, the decoding algorithm recovers the original data by finding the most probable path swept by the encoder FSM. 
This is done by investigating the likelihood of many different paths. 
Since the number of possible paths grow exponentially, the Viterbi decoding algorithm employs a dynamic programming approach to efficiently find the most probable path as the following \cite{Viterbi}. 

First, for branch $ij$ from state $i$ to state $j$ at stage $t \in [0,\N)$, \textbf{branch metric} $\BM^{ij}_{t}$ is computed as 
\begin{equation}
\BM^{ij}_t = \sum_{b=0}^{\B-1} (-1)^{\displaystyle \BO^{ij}[b]} \times \llr_t[b]
\label{eq:BM}
\end{equation}
where, $\BO^{ij}[b]$ and $\llr_t[b]$ denote bit $b$ of $\BO^{ij}$ and $\llr_t$, respectively. $b \in [0,\B)$. The term $\llr_t$ denotes the received $\B$ bits at the decoder input at time $t$. Basically, $\BM^{ij}_t$ indicates the amount of similarity between the received data and the output of branch $ij$. 
%
%
%
Next, for state $j$ at stage $t$, \textbf{path metric} $\PM^j_t$ is computed as 
\begin{equation}
\PM^j_t = \max_{i \in \text{prv}(j)} \big( 
\PM^{i}_{t-1} + \BM^{ij}_t  
\big)
\label{eq:PM}
\end{equation}
where, $i$ iterates over the previous states of state $j$ in the FSM. Note that every state has two previous states, i.e., two input branches. 
The above equation is known as ACS (add, compare and select) operation. Basically, $\PM^j_t$ is formed by accumulating a series of branch metrics which eventually end at state $j$ at stage $t$, and also, have the highest possibility. In other words, $\PM^j_t$ indicates the likelihood of the most probable path which ends at state $j$ till stage $t$.

While computing $\PM^j_t$, the selected previous state $\SP^j_t$ which maximized $\PM^j_t$ needs to be saved as well. It basically indicates one stage of the \textbf{survivor path} which ends at state $j$ at stage $t$. 
\begin{equation}
\SP^j_t = \argmax_{i \in \text{prv}(j)} \big( 
\PM^{i}_{t-1} + \BM^{ij}_t  
\big)
\label{eq:SP0}
\end{equation}
%

The above calculations constitute the forward procedure in the Viterbi decoding algorithm. See Alg.~\ref{alg:serialVit1}. The outer loop iterates through all stages $t \in [0, \N)$ and the inner loop iterates through all states $j \in [0, 2^{\K-1})$.

Once the forward procedure is complete, the backward procedure is performed as the following. 
First, the most probable survivor path at the last stage, i.e., $t=\N-1$, is selected as the winner, and that specific path is traced back to the first stage, i.e., $t=0$. At every stage during the trace-back, the decoder output (which is ideally equal to the original data) is formed based on $\BI^{ij}$. The two operations are called \textbf{trace-back} and \textbf{decoding}. See Alg.~\ref{alg:serialVit2}.

\begin{algorithm}[tp]
	\begin{algorithmic}[1]
		\Input $\llr [\B][\N]$ 
		\Output $\SP[2^{\K-1}][\N]$
		
		\State Initialize $\PM[2^{\K-1}][\N]$ to zero
		\For{$t=0$ to $\N-1$}
		\For{$j=0$ to $2^{\K-1}-1$}
		\State Set $i'$ and $i''$ equal to the two previous states of $j$
		\State Compute $\BM^{i'j}_{t}$ and $\BM^{i''j}_{t}$ according to \eqref{eq:BM}
		\State $\PM' = \PM[i',t-1] + \BM^{i'j}_{t}$
		\State $\PM'' = \PM[i'',t-1] + \BM^{i''j}_{t}$
		\If{$\PM' > \PM''$}
		\State $\PM[j,t] = \PM'$
		\State $\SP[j,t] = i'$
		\Else
		\State $\PM[j,t] = \PM''$
		\State $\SP[j,t] = i''$
		\EndIf
		\EndFor
		\EndFor
	\end{algorithmic}
	\caption{\small The first step, i.e., the forward procedure, in the Viterbi decoding method.}
	\label{alg:serialVit1}
\end{algorithm}

\begin{algorithm}[tp]
	\begin{algorithmic}[1]
		\Input $\SP[2^{\K-1}][\N]$
		\Output $Out[\N]$ 
		\State  $j^* = \argmax\limits_{j \in [0,2^{\K-1})} \PM[j,\N-1]$
		\For{$t=\N-1$ to $0$}
		\State $i = \SP[j^*,t]$
		\State $Out[t] = \BI^{ij^*}$ 
		\State $j^*=i$
		\EndFor
	\end{algorithmic}
	\caption{\small The second step, i.e., the backward procedure, in the Viterbi decoding method.}
	\label{alg:serialVit2}
\end{algorithm}

\begin{table*}[tp]
	\centering
	\includegraphics[width=0.6\textwidth]{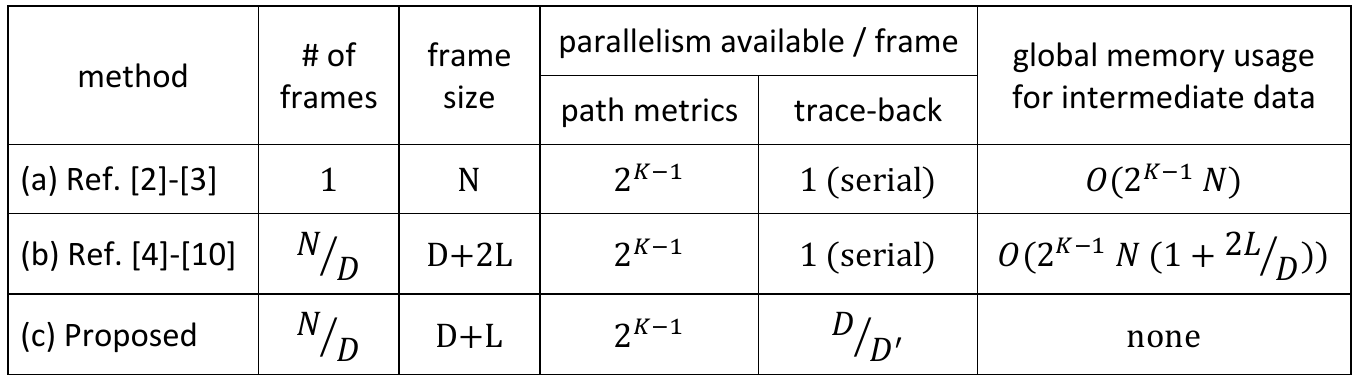} 
	\caption{Comparing the proposed method with previous works, in terms of the amount of parallelism made available, and the amount of global memory usage for intermediate data.}
	\label{fig:comparetable}
\end{table*}

\subsection{Hard-decision vs. Soft-decision} 

The input to the Viterbi decoding algorithm can be represented in either hard-decision mode or soft-decision mode. In  \textbf{hard-decision} mode, every bit is simply represented by either zero or one. 
In \textbf{soft-decision} mode, however, every bit in the input of the decoder is a \textbf{log likelihood ratio (LLR)} that is formed by the receiver circuit based on the probability that the received bit is zero or one. A larger positive LLR means a larger probability of zero, and a larger negative LLR means a larger probability of one. 
Basically, in addition to the indication of the value of the input bit, the truth of this indication is provided as well.

The Viterbi decoder can take advantage of this additional information in order to better recover the original data. Bit error rate (BER) is lower in the soft-decision mode by about $2.3$~dB. This comes at the cost of higher computational requirement.

\subsection{CUDA Parallel Programming API}
\label{sec:prelim:cuda}

CUDA is a parallel programming API for Nvidia GPUs. GPU is a massively parallel processor with hundreds to thousands of cores. 
CUDA follows a hierarchical programming model. At the top level, computationally intensive functions are specified by the programmer as CUDA \textbf{kernels}. A kernel is specified as a sequential function for a single \textbf{thread}. The kernel is then launched for parallel execution on the GPU by specifying the number of concurrent threads. 

Threads are grouped into \textbf{blocks}. A kernel consists of a number of blocks, and every block consists of a number of threads. 
%
%
In order to identify blocks within a kernel, and also, threads within a block, a set of indices are used in the CUDA API, for instance,  $blockIdx.x$ as the block index in dimension $x$ within a kernel, and $threadIdx.x$ as the thread index in dimension $x$ within a block. 

\section{Previous GPU-Accelerated Viterbi Decoder Methods}
\label{sec:relwork}

\begin{figure}[tp]
	\centering
	\includegraphics[width=0.9\columnwidth]{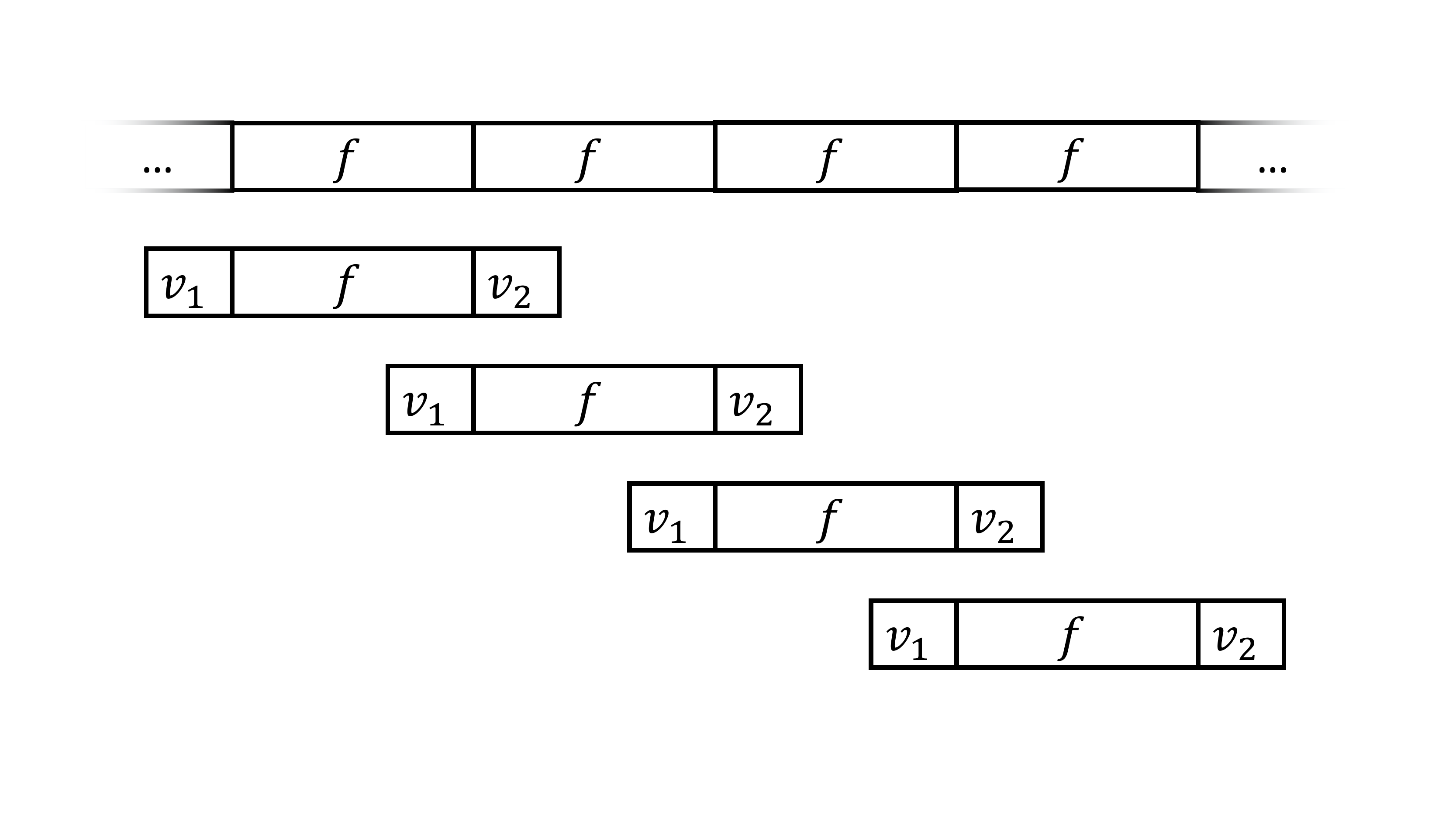} 
	\caption{framing method to make the algorithm parallel}
	\label{fig:relworks}
\end{figure}

The Viterbi decoder algorithm is inherently a sequential procedure and the amount of available parallelism is minimal. 
In specific, calculating the branch metrics is the only step which can be fully parallelized, and path metric calculations can only be partially parallelized using at most $2^{\K-1}$ threads, e.g., $2^{7-1}=64$ threads, where each thread sequentially iterates over $\N$ stages. This approach was proposed in \cite{zhang2009,kim2010ieeecomm}. See Fig.~\ref{fig:relworks}(a). 
%
%


In order to better utilize the parallel computing capabilities of GPU and increase the throughput, the decoder output can be estimated by dividing the $\N$ stages into many small frames (tiles) of $\F$ stages each \cite{lin2011tiling, gautam2014opencl, lee2013, lee2014}. Every frame is processed in parallel with other frames, and decodes $\F$ out of $\N$ output bits. 
This tiling scheme increases the amount of available parallelism by a factor of $\nicefrac{\N}{\F}$. However, BER is degraded because not all previous history is available in a frame. In order to reduce BER degradation, consecutive frames should have small overlaps in order to carry enough history for correct decoding \cite{lin2011tiling, gautam2014opencl, lee2013, lee2014}. 
As shown in Fig.~\ref{fig:relworks}(b), every frame has an overlap of length $\Ovr$ with its neighbor frames, that is composed of left overlap $\Ovr_1$ and right overlap $\Ovr_2$. Hence, to generate $\F$ decoded bits as the output, a frame needs to process $\F+\Ovr$ stages of the original Viterbi decoding algorithm. 


In the forward procedure, the resulting survivor paths need to be stored in GPU global memory for later use in the backward procedure \cite{lin2011tiling,gautam2014opencl,lee2013,lee2014}. The amount of GPU global memory required to store the survivor paths is in the order of 
\begin{equation}
O \Big( 2^{\K-1} \times \N \times (1+\frac{\Ovr}{\F}) \Big)
\end{equation}
because there are $\nicefrac{\N}{\F}$ frames, and every frame requires $O(2^{\K-1} \times (\Ovr+\F))$ space from GPU global memory. 
The parallel algorithms proposed in \cite{li2013,li2014} improved the throughput by judiciously combining the execution of multiple frames in order to coalesce the memory accesses of the survivor paths in global memory. The coalesced memory accesses result in higher throughput.

In \cite{peng2016}, in addition to the tiling scheme and coalescing accesses of survivor paths, branch metrics are efficiently computed according to specific repetitive patterns which help to share computations. In addition, data transfers between CPU and GPU are optimized, in specific, by employing multiple CUDA streams. 


%

\section{Proposed Parallel Algorithm}
\label{sec:alg}

This section presents our proposed parallel algorithm for execution of the Viterbi decoding algorithm on GPU based on CUDA framework. 

\subsection{Unified Kernel}
\label{sec:alg:unified}

As discussed above, the best previous parallel algorithms coalesce access to survivor paths in GPU global memory in order to minimize the associated latency and increase the throughput \cite{li2013, li2014, peng2016}. 
We propose to take a radically different approach, which I) highly reduces this memory access latency, and  II) exposes a higher degree of parallelism. 

In specific, we propose to parallelize the backward procedure (i.e., trace-back and decode) which enables us to efficiently combine both the forward procedure (i.e., branch metric, path metric and survivor path computations) and the backward procedure into a single unified parallel kernel. 
This unified kernel, in turn, allows the survivor paths to be stored in GPU \textbf{shared memory}, and therefore, global memory access is no longer required for storing and retrieving this intermediate data. 
As a result, the associated access latency to GPU global memory for survivor paths is removed completely and throughput is increased. 
In CUDA programming API, every block has access to a small, on-chip and low-latency memory, called shared memory. The shared memory of a block is accessible to all threads within that block, but not to any thread from other blocks.

It is very important to note that once the execution of a parallel kernel is completed, its shared memory data is no longer available, for instance, to subsequent kernels. Therefore, survivor paths cannot be stored in shared memory in any of the previous solutions, because the forward and backward procedures are performed in separate kernels \cite{zhang2009, kim2010ieeecomm, lin2011tiling, gautam2014opencl, lee2013, lee2014, li2013, li2014, peng2016}. 

The reason previous solutions had to employ separate kernels is that the forward and backward procedures in the Viterbi decoding algorithm have very different degrees of parallelism. In particular, in the forward procedure, branch metric computations are fully parallelizable, and the degree of parallelism available for computing the path metrics is $2^{\K-1}$. At the other hand, the backward procedure is completely serial. See the first row in Table \ref{fig:comparetable}. Even after applying the tiling scheme, the amount of parallelism made available in the backward procedure is minimal. In specific, it is equal to the number of frames, because the backward procedure is still performed sequentially in every frame \cite{lin2011tiling, gautam2014opencl, lee2013, lee2014, li2013, li2014, peng2016}. 
In the proposed solution, however, the backward procedure is parallelized. See the last row in Table \ref{fig:comparetable}.

A parallelized methods for the backward procedure are presented in Sections \ref{sec:alg:pt1}. However, before discussing the parallelized backward procedure, the proposed ideas in optimizing branch metric and path metric computations are discussed in Sections \ref{sec:alg:bm} and \ref{sec:alg:pm}.

\begin{algorithm}[tp]
	\begin{algorithmic}[1]
		\Input $\llr [\B][\N]$ 
		\Output $Out[\N]$ 
		\Block $\frac{\N}{\F}$
		\Thread $2^{\K-1}$
		
		\State \_\_shared $\BM [ 2^{\K-1}][\F+\Ovr][2]$ 
		\State \_\_shared $\PM [ 2^{\K-1}][\F+\Ovr]$ 
		\State \_\_shared $\SP [ 2^{\K-1}][\F+\Ovr]$ 
		\For{$t=0$ to $\F+\Ovr-1$}
			\State $j=threadIdx.x$
			\State Set $i'$ and $i''$ equal to the two previous states of $j$
			\State Compute $\BM[j][t][0]=\BM^{i'j}_{t}$ and $\BM[j][t][1]=\BM^{i''j}_{t}$
			\State $\PM' = \PM[i'][t-1] + \BM[j][t][0]$
			\State $\PM'' = \PM[i''][t-1] + \BM[j][t][1]$	
			\State \_\_syncthreads(~)
			\If{$\PM' > \PM''$}
				\State $\PM[j][t] = \PM'$
				\State $\SP[j][t] = i'$
			\Else
				\State $\PM[j][t] = \PM''$
				\State $\SP[j][t] = i''$
			\EndIf
		\EndFor

		\State  $j^* = \argmax\limits_{j \in [0,2^{\K-1})} \PM[j][\N-1]$
		\For{$t=\F+\Ovr-1$ to $0$}
			\State $i = \SP[j^*][t]$
			\State $out[t] = \BI^{ij^*}$ 
			\State $j^*=i$
		\EndFor
	\end{algorithmic}
	\caption{The forward and backward procedures in a unified parallel kernel.}
	\label{alg:unified}
\end{algorithm}

\begin{figure}[tp]
	\centering
	\includegraphics[width=0.9\columnwidth]{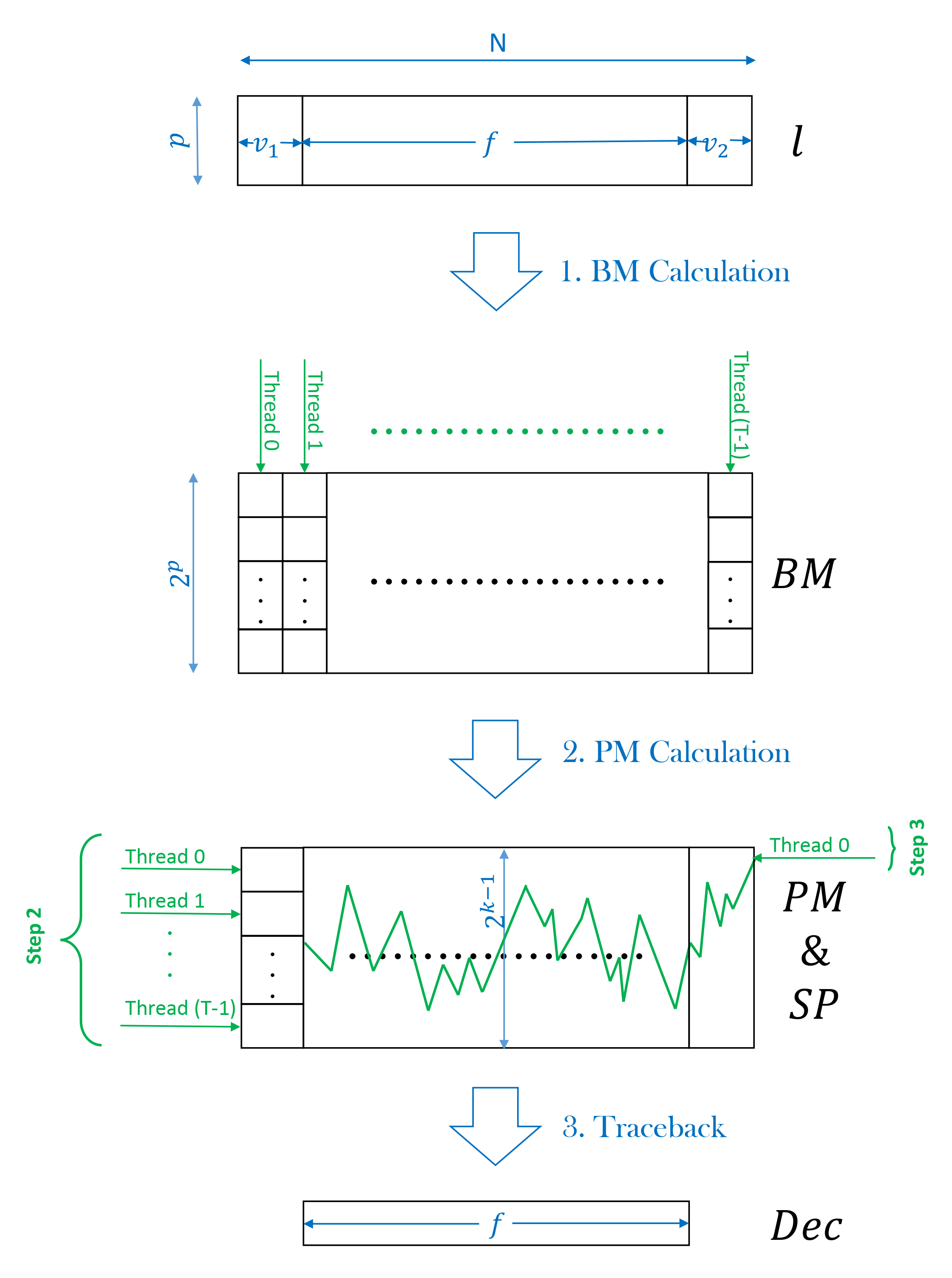} 
	\caption{Viterbi algorithm steps in each block of GPU}
	\label{fig:process}
\end{figure}

\subsection{Branch Metrics}
\label{sec:alg:bm}

\begin{figure}[tp]
	\centering
	\includegraphics[width=0.9\columnwidth]{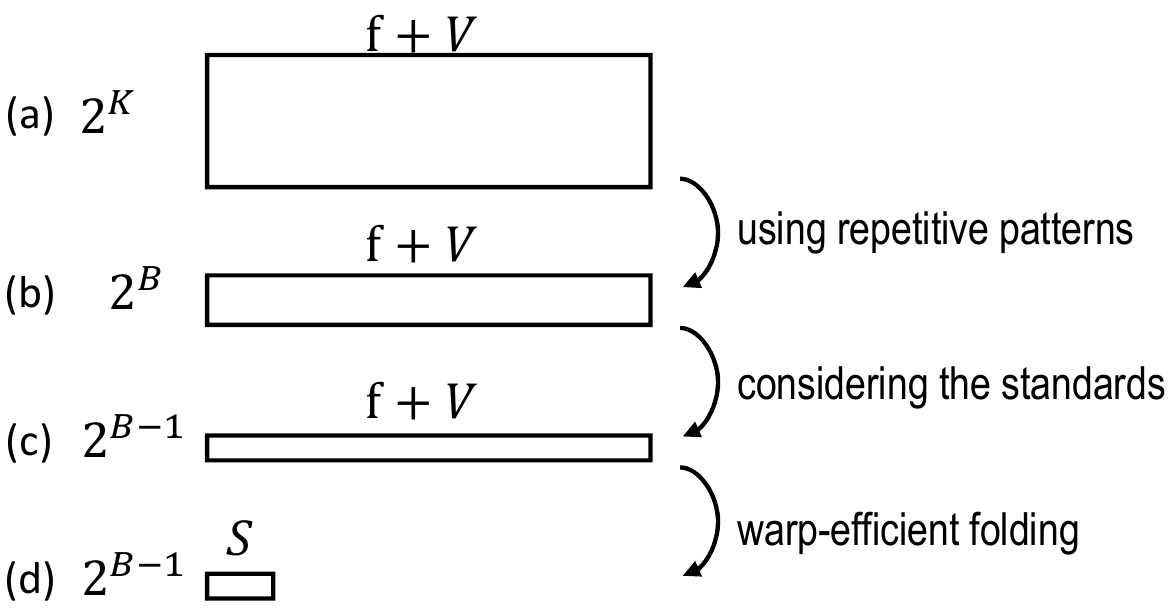} 
	\caption{Optimizing the memory requirement for storing branch metrics in shared memory.}
	\label{fig:bm_opt}
\end{figure}

In our proposed unified-kernel approach, every frame is processed in parallel to other frames, and decodes $\F$ out of $\N$ output bits. Every frame involves $\F+\Ovr$ stages, and every stage has $2^{\K}$ states. In the simplest form, every block processes one frame using $2^{\K-1}$ threads. Hence, $2^{\K} \times (\F+\Ovr)$ branch metrics need to be computed in every block. As shown in Fig.~\ref{fig:bm_opt}(a), storing all such values in shared memory requires a two-dimensional matrix of size 
\begin{equation}
2^{\K} \times (\F+\Ovr),
\label{eq:shmem:bm_a}
\end{equation}
which is not an efficient use of shared memory. 
GPU has a hardware scheduler that automatically assigns multiple blocks to every streaming multi-processor (SM). Since the amount of shared memory in every SM is limited, the smaller the shared-memory usage of every block, the larger the number of blocks that are assigned to every SM, and hence, the higher the achieved throughput. 
We employ the following strategies in order to reduce the amount of shared memory usage in every block.

\vskip 2mm
\textit{On-the-fly Computation:} During the computation of the path metrics, branch metrics can be computed on-the-fly. In other words, equations \eqref{eq:BM} and \eqref{eq:PM} can be evaluated at the same time. This approach does not require storing any of the branch metrics in shared memory.

\vskip 2mm
\textit{Repetitive Patterns:} The above approach leads to many redundant computations. This is because branch metrics follow specific repetitive patterns \cite{peng2016}. See \eqref{eq:BM}. In every stage $t$, $\llr_t$ is constant. Hence, since $\BO^{ij}$ has $\B$ bits, there are only $2^\B$ unique branch metric values in every stage \cite{peng2016}. Therefore, a more optimized approach is to first compute 
\begin{equation}
2^\B \times (\F+\Ovr) 
\label{eq:shmem:bm_b}
\end{equation}
unique branch metrics and store them in shared memory, then compute the path metrics. 

\vskip 2mm
\textit{Standard Convolutional Codes:} We can cut the above shared-memory requirement for branch metrics to half. Since $\overline{\BO^{ij}} = 2^\B - \BO^{ij}$, we have 
\begin{align}
\sum_{b=0}^{\B-1} (-1)^{\displaystyle \overline{\BO^{ij}[b]}} \times \llr_t[b] = -\sum_{b=0}^{\B-1} (-1)^{\displaystyle \BO^{ij}[b]} \times \llr_t[b]
\label{eq:BMSymmetric}
\end{align}
which means half of the branch metric values are complements of the other half. As a result, a matrix of size 
\begin{equation}
2^{\B-1} \times (\F+\Ovr)
\end{equation}
is sufficient to store the branch metrics. 

\vskip 2mm
\textit{Wrap-Efficient Sub-folding:} In above approaches, branch metrics for whole the stages are evaluated at first and then path metrics calculation are performed. These two loops that have $\F+\Ovr$ iterations can be folded and performed in $S$ iterations if both of them are in an outer loop of size $\nicefrac{\F+\Ovr}{S}$. An important point is that $S$ should be selected wisely to avoid violating considerations about wrap efficiency.

\subsection{Path metrics}
\label{sec:alg:pm}

Similar to the branch metrics, $2^{\K-1} \times (\F+\Ovr)$ path metrics need to be computed in every block, and storing all such values in shared memory is not efficient. 
As shown in \eqref{eq:PM}, path metrics $\PM$ are computed iteratively. In other words, to compute path metrics in stage $t$, it is enough to have access to path metrics only from the previous stage. Hence, with proper thread synchronizations from stage $t-1$ to stage $t$, an array of size 
%
$2^{\K-1}$ 
%
is sufficient to store the path metrics in shared memory.

\subsection{Parallel Traceback Algorithm}
\label{sec:alg:pt1}

The traceback step is intrinsically serial and should be done by one thread. However, it leads to a decrease in GPU utilization as other threads are idle. So a parallel traceback algorithm should be provided to solve this problem. In order to parallelize the traceback step, two different algorithms are provided in this article.

\begin{figure}[t]
	\centering
	\includegraphics[width=0.9\columnwidth]{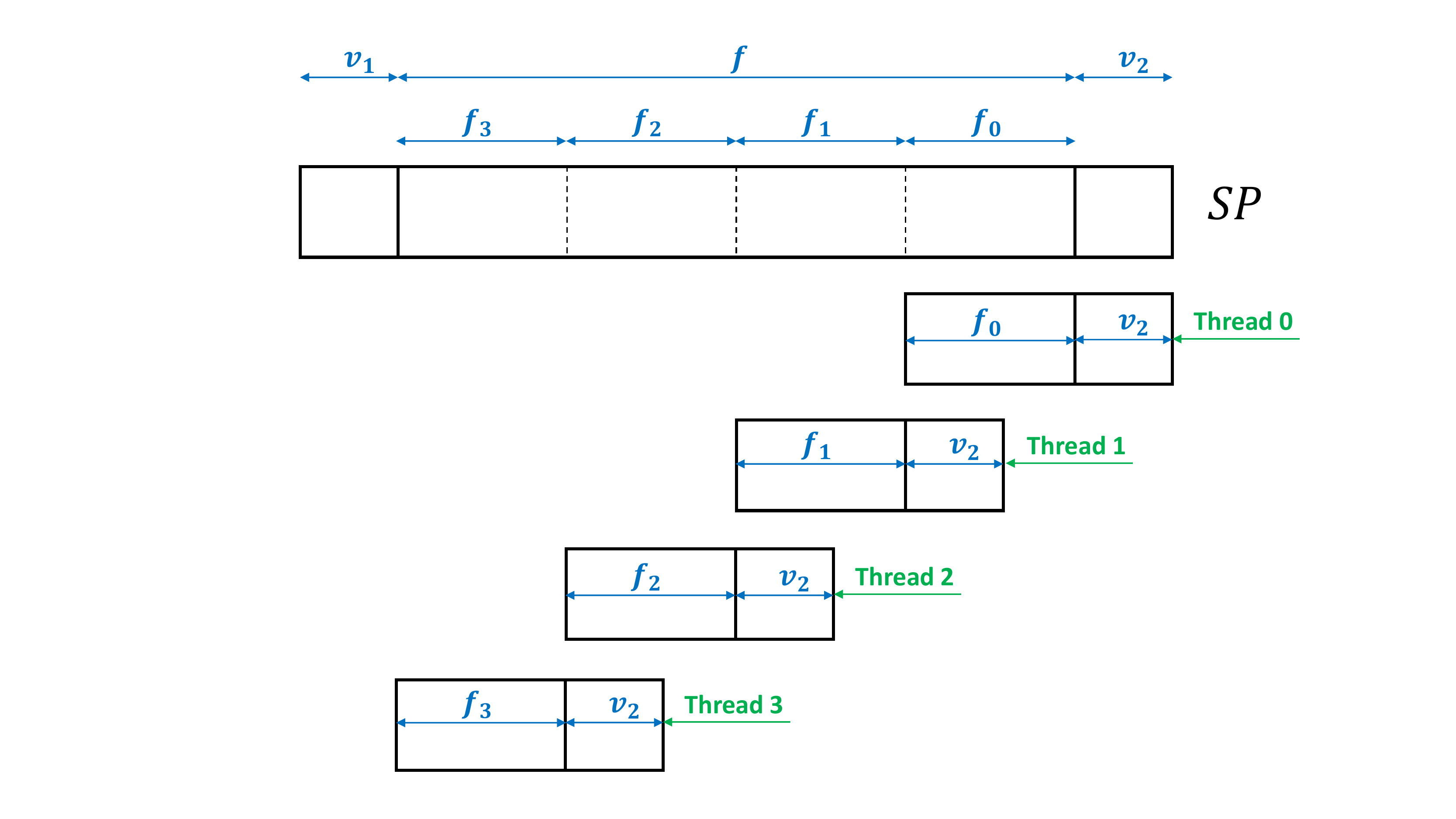} 
	\caption{Parallel traceback algorithm with 4 threads}
	\label{fig:pt1}
\end{figure}

In the first parallel traceback, the non-overlapping part of the frame should be divided into some subframes that every subframe is decoded by one thread in parallel. However, in the Viterbi algorithm, at the point that traceback starts to some stages later, decoded bits are not reliable and mustn't be stored. After some stages, the survivor path that is traced back converges to its correct states and decoded bits become reliable and can be stored. that is the reason that the main frame should overlap its right-handed neighbor and the decoded bits of the overlapping part are not stored and is only used for the convergence of the survivor path. The same approach should be taken in the parallel traceback. Each subframe should overlap its right-handed neighbor as depicted in Fig.~\ref{fig:pt1} and the decoded bits of the overlapping part is not stored. The length of the overlapping part depends on the coding parameters. Thus, it can be the same as the main frame like Fig.~\ref{fig:pt1}. 

Another problem to be solved is that the starting state of the traceback is the state that its final path metric is the maximum. So in this parallel traceback algorithm, the final path metrics of all subframes that are the path metrics of the stages of subframes right boundaries are needed. However, when the traceback step is going to start, only the path metrics of the final stage is available. To handle this problem, the starting point can be random. As a result, the convergence will take longer. So the overlapping length $\Ovr_2$ should be increased. Another approach is that, instead of storing all path metrics of intermediate stages, only the state with maximum path metric can be obtained and stored. Thus, a reasonable amount of memory is used and convergence is not postponed.

\subsection{De-puncturing}
\label{sec:alg:depunc}

In the primary form of the convolutional encoder, the rate of the coding can be only $1/a$ where $a$ is a natural number greater than $2$. Puncturing is the elimination of some bits from the output of the encoder to produce codes with a wider variety of rates. This elimination should be done using a standard pattern. The standard pattern is like a mask with a fixed size that drops some bits and is replicated on the data. As the number of the output bits decreases after puncturing, the rate of the encoder increases. Since some information is removed, BER increases after puncturing. In the receiver, eliminated bits is replaced with zeros to make them neutral to metrics. It is called depuncturing. Then the Viterbi algorithm is applied like before.

In the implementation on GPU, a depuncturing step should be added at the beginning of the algorithm before branch metric calculation. In order to parallelize the step, each thread can depuncture some bits independently. Additionally, another array should be allocated to store depunctured data. As explained before, it is better to store this intermediate data in the shared memory.

In order to avoid the divergence of the blocks, all frames should start at the beginning of a pattern mask both in their overlapping and non-overlapping part. It means that $\F$, $\Ovr_1$ and $\Ovr_2$ should be a multiple of the size of the patter mask.

\subsection{Memory Management}
\label{sec:alg:mm}

\begin{figure}[t]
	\centering
	\includegraphics[width=0.9\columnwidth]{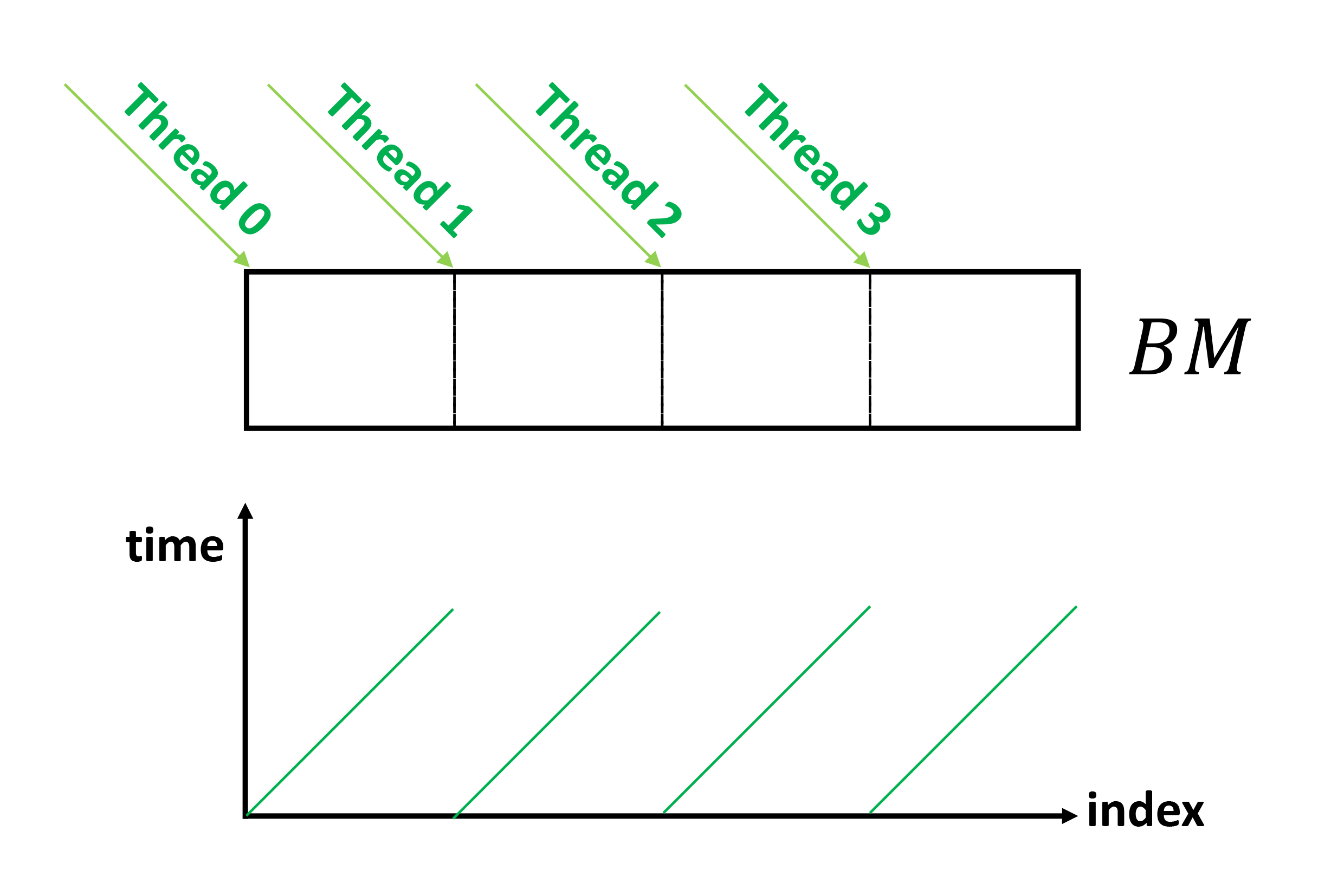} 
	\caption{Branch metric computation.}
	\label{fig:memoryNotCoa}
\end{figure}

\begin{figure}[t]
	\centering
	\includegraphics[width=0.9\columnwidth]{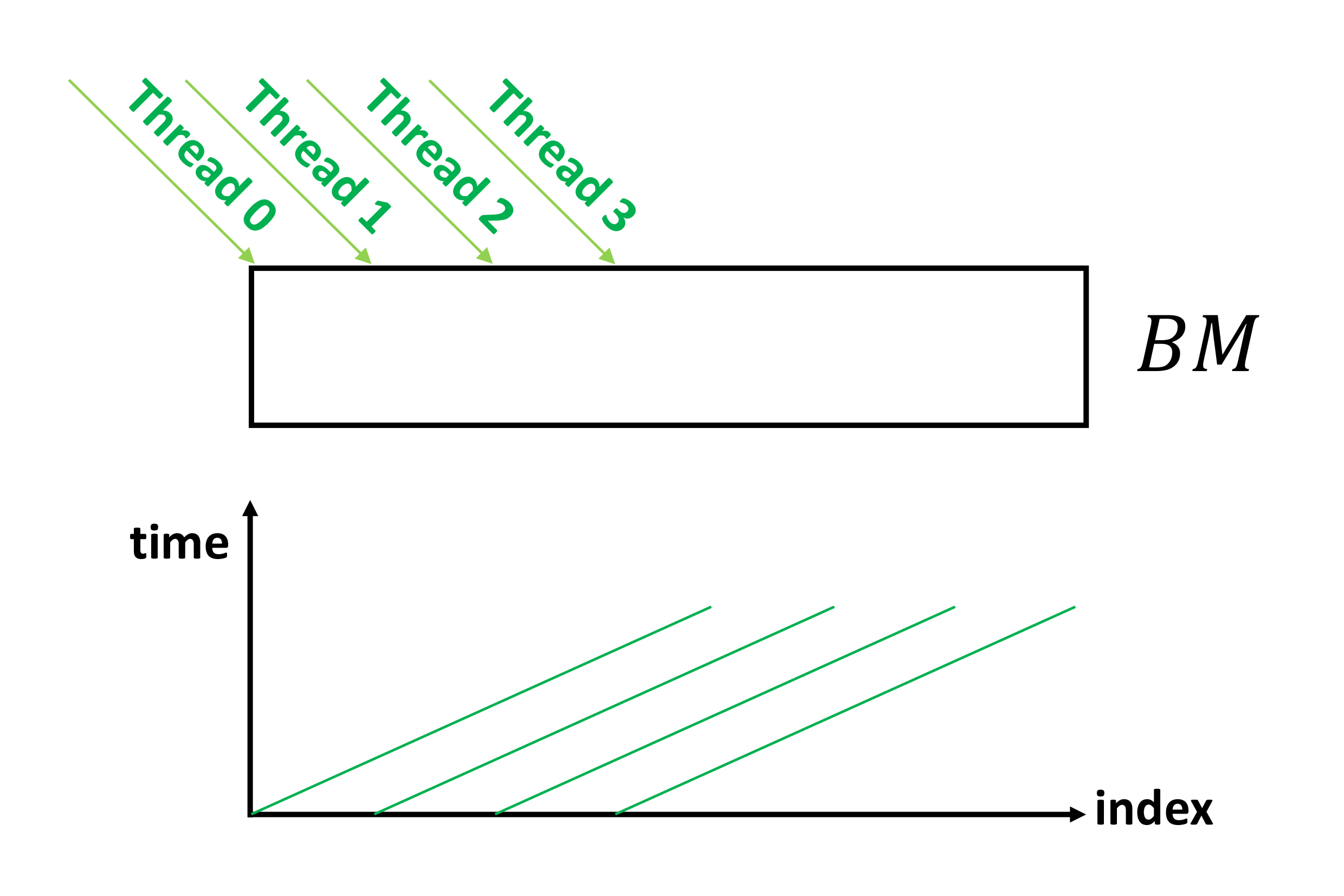} 
	\caption{Coalesced branch metric computation.}
	\label{fig:memoryCoa}
\end{figure}

The important parameter of memory efficiency in GPU is coalesced access. In the branch metric calculation step, the $BM$ array is divided into some subarrays and each subarray is calculated by one independent thread as depicted in Fig.~\ref{fig:memoryNotCoa}. In such an approach, distant elements are accessed at a time as the graph in Fig.~\ref{fig:memoryNotCoa} illustrates. Thus, it is better that all threads fill $T$ successive columns of the array where $T$ is the number of the threads. Then all threads calculate the next $T$ columns as depicted in Fig.~\ref{fig:memoryCoa}. As a result, threads' access is coalesced. This approach can be employed also in the filling of the depunctured frame array.

Another modification that can lead to coalesced access is handling more than one frame in each block. Even though GPU analyses the kernel to decide how many blocks can be assigned to each SM, handling it in the kernel can be effective. when some frames are running on one SM under the decision of GPU, the amount of shared memory block needed is allocated after each other. As all blocks are running the same algorithm, their shared memory blocks are accessed in the same part simultaneously. As a result, the arrays that are accessed simultaneously by different GPU blocks of the same SM are distant as much as the size of the shared memory. However, when several frames are handled in a block, their shared memory block can be interleaved, meaning arrays size is considered for all frames and the arrays are consecutive not whole the shared memory. Therefore, memory cells accessed at the same time will closer to each other.

Another aspect of memory optimization is the size of the shared memory used. In this algorithm, according to the report of Visual Profiler, shared memory is the bottleneck and reduction of the size of the used shared memory can be helpful. An approach used to do so is reusing shared memory. There are three steps in this algorithm and each step has some arrays stored in shared memory. However, each of these intermediate arrays is used in a limited number of steps and their life time is not whole the kernel life time. Therefore, if there are two arrays needed in two distinct life time, both arrays can be stored in the same place. It is important to keep in mind that if their types are different, the pointer alignment should be satisfied for the larger type. As a result, the size of the used shared memory will be reduced and GPU utilization will enhance. In this implementation, $SP$ and depunctured frame are two arrays that can use the same memory. boundary states of the second parallel traceback algorithm and $PM$ can also be stored in the same place.

\section{Experimental Evaluation}

\subsection{Setup}

The hardware platform is a server with an Intel Xeon CPU operating at $2.5$~GHz. We experiment with Tesla V100 GPU.
We employ Ubuntu OS $18.04$ and CUDA version $10.2$.

We experiment with a widely-used standard convolutional code, namely, $(2,1,7)$, i.e., code rate $\nicefrac{1}{2}$ and constraint length $7$, with generator polynomials $171$ and $133$. This configuration is shown in Fig.~\ref{fig:encoder}. In addition, we also consider the puncturing rates of $\nicefrac{2}{3}$ and $\nicefrac{3}{4}$. 

\subsection{BER Performance}

\begin{figure}[t]
	\centering
	\includegraphics[width=0.9\columnwidth]{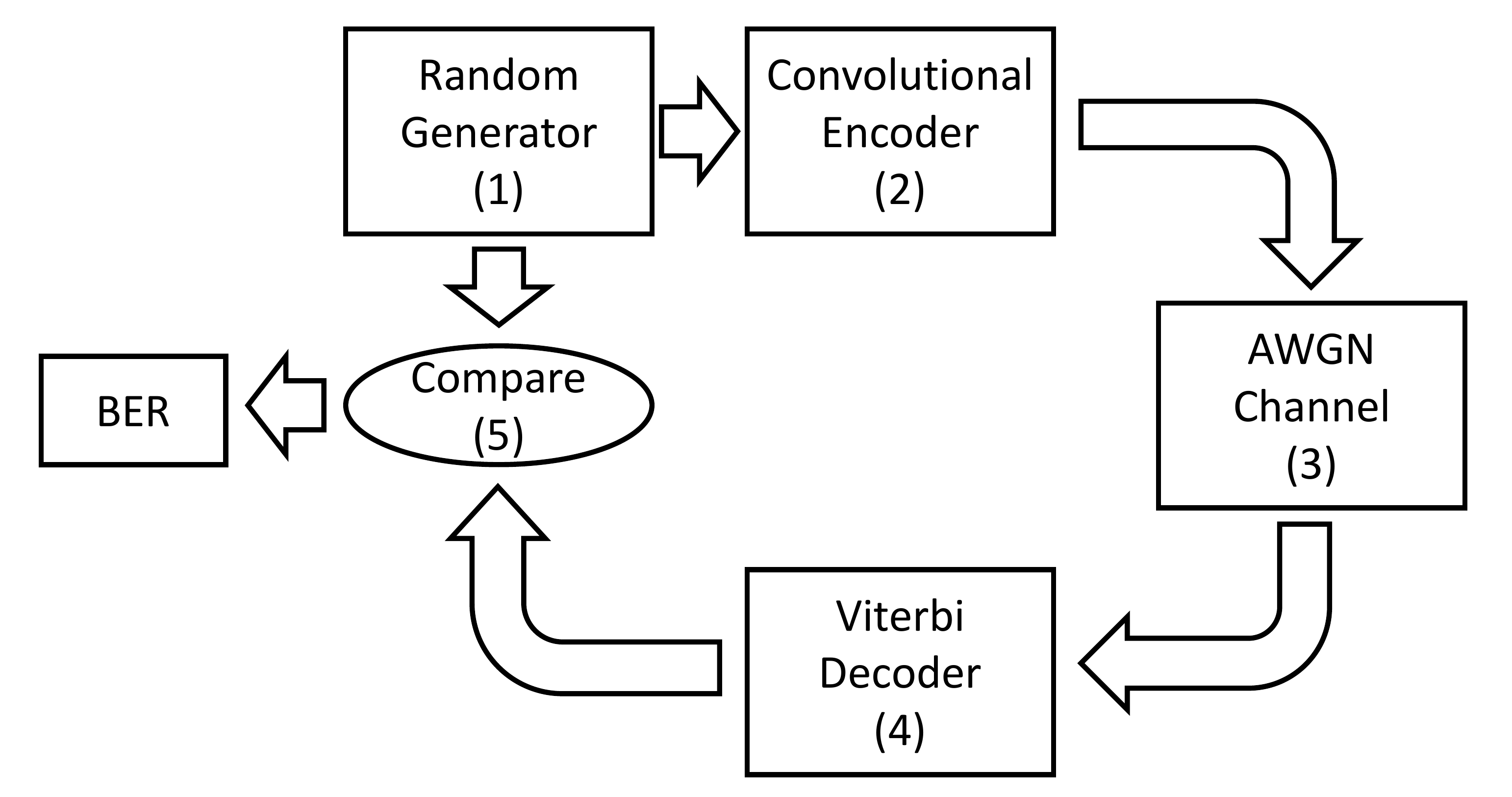} 
	\caption{The block diagram of the verification system}
	\label{fig:verifSys}
\end{figure}

In order to verify the implementation, the system shown in Fig.~\ref{fig:verifSys} is employed. At the first step, a vector of uniformly distributed bits is generated and, at step $2$, passed to convolutional encoder. This part is the simulated transmitter. Then, at step $3$, encoded bits are transmitted in an AWGN channel with a specific $E_b/N_0$. Assuming that the BPSK modulation is used, the channel simulation is done by adding a vector of normally distributed values with standard deviation of $2^{\nicefrac{-(E_b/N_0)}{20}}$. Having generated a noisy coded vector, at step $4$, the simulated receiver can decode the signal and produce an output vector. At last, comparing the decoder output with the bits generated at the first step, Bit Error Rate (BER) will be obtained. It should be also noted that the BER value is reliable if enough data is generated and tested in the verification system. As a rule of thumb, if a vector of size $n$ is generated in the first step, only the BER value more than $\frac{100}{n}$ will be valid.

The process shown in Fig.~\ref{fig:verifSys} can produce the BER for a specific $E_b/N_0$. Complete verification is done by drawing the BER curve over a range of $E_b/N_0$ values and comparing it with the theoretical one that can be generated by MATLAB BER tool which is invoked by "bertool" command. Implementation parameters can be tuned this way.

\begin{figure}[t]
	\centering
	\includegraphics[width=0.9\columnwidth]{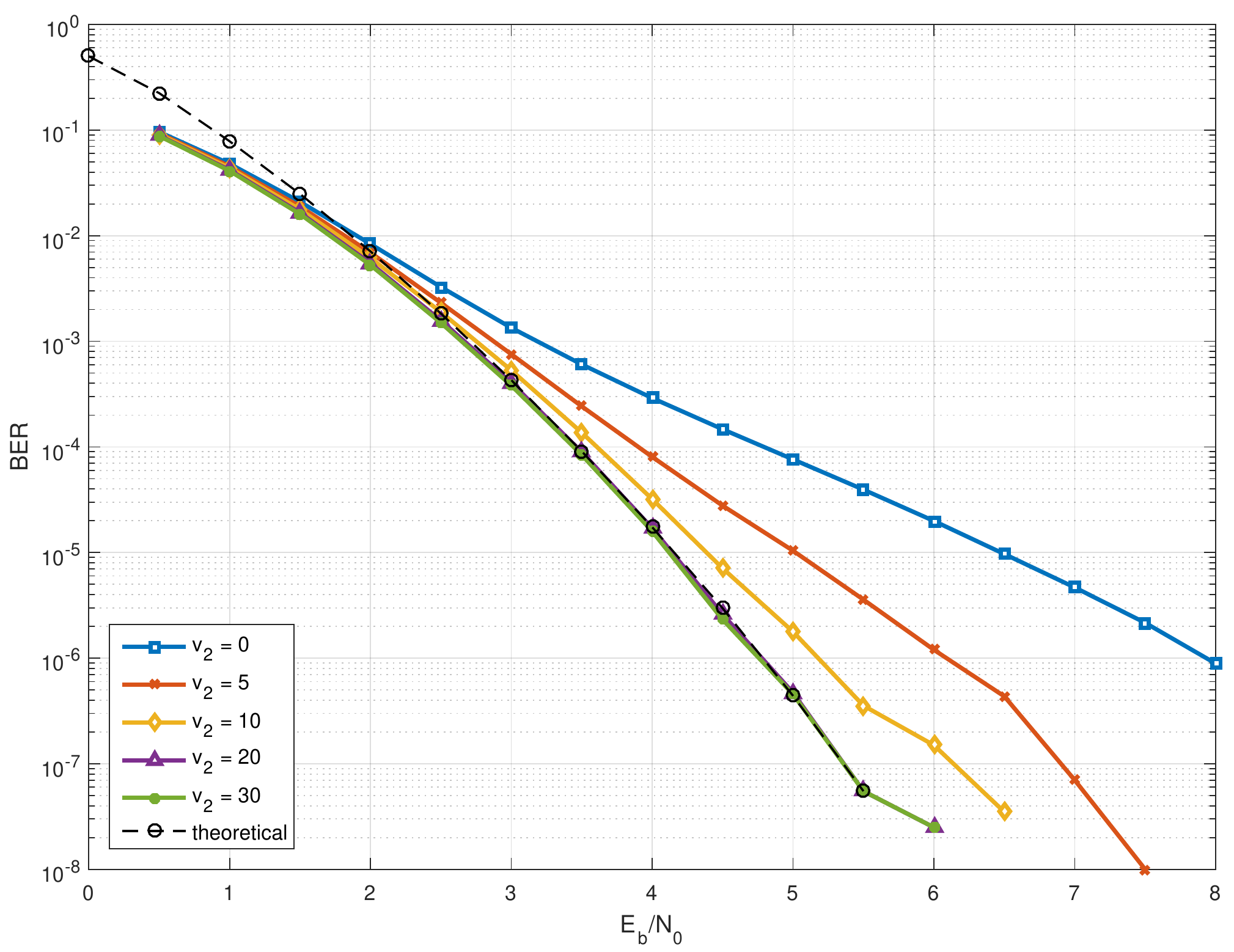} 
	\caption{The effect of $v_2$ on BER for $v_2=20$ and $f=256$}
	\label{fig:ber:v2}
\end{figure}

As illustrated in Fig.~\ref{fig:relworks}, input vector is divided into overlapping blocks. In this approach, there are three parameters, namely, $f$, $v_1$, and $v_2$. Experiments has shown that $v_1$ has almost nothing to do with BER and the effect of $f$ is negligible. By contrast the effect of $v_2$ is considerable since it determines the convergence in traceback. Fig.~\ref{fig:ber:v2} shows that for $v_2=20$, theoretical performance is achieve and no considerable improvement can be gained by $v_2>20$.

A metric can be defined to show the code performance for every set of parameters. What concerns is the distance between practical and theoretical curve. Therefore, the metric should represents this distance. The distance in the dimension of $E_b/N_0$ is reasonable since it shows that in order to achieve a specific BER, how much clearer the signal should be than it should be in theory. Table.~\ref{table:ber} shows this metric for a wider range of $f$ and $v_2$ than Fig.~\ref{fig:ber:v2}.

\begin{table}[h]
	\begin{center}
		\caption{The effect of $f$ and $v_2$ on BER}
		\label{table:ber}
		\begin{tabular}{|l||*{5}{l|}}
			\hline
			\backslashbox{$v_2$}{$f$} & \textbf{32} & \textbf{64} & \textbf{128} & \textbf{256} & \textbf{512} \\
			\hline
			\hline
			\textbf{10} & $0.72$ & $0.48$ &  $0.31$ & $0.18$ & $0.12$ \\
			\hline	
			\textbf{20} & $0.15$ & $0.090$ &  $0.044$ & $0.040$ & $0.039$ \\
			\hline
			\textbf{30} & $0.030$ & $0.016$ &  $0.0069$ & $0.022$ & $0.033$ \\
			\hline
			\textbf{40} & $0.0040$ & $0.00097$ &  $0.0032$ & $0.025$ & $0.034$ \\
			\hline
		\end{tabular}
	\end{center}
\end{table} 

\begin{figure}[t]
	\centering
	\includegraphics[width=0.9\columnwidth]{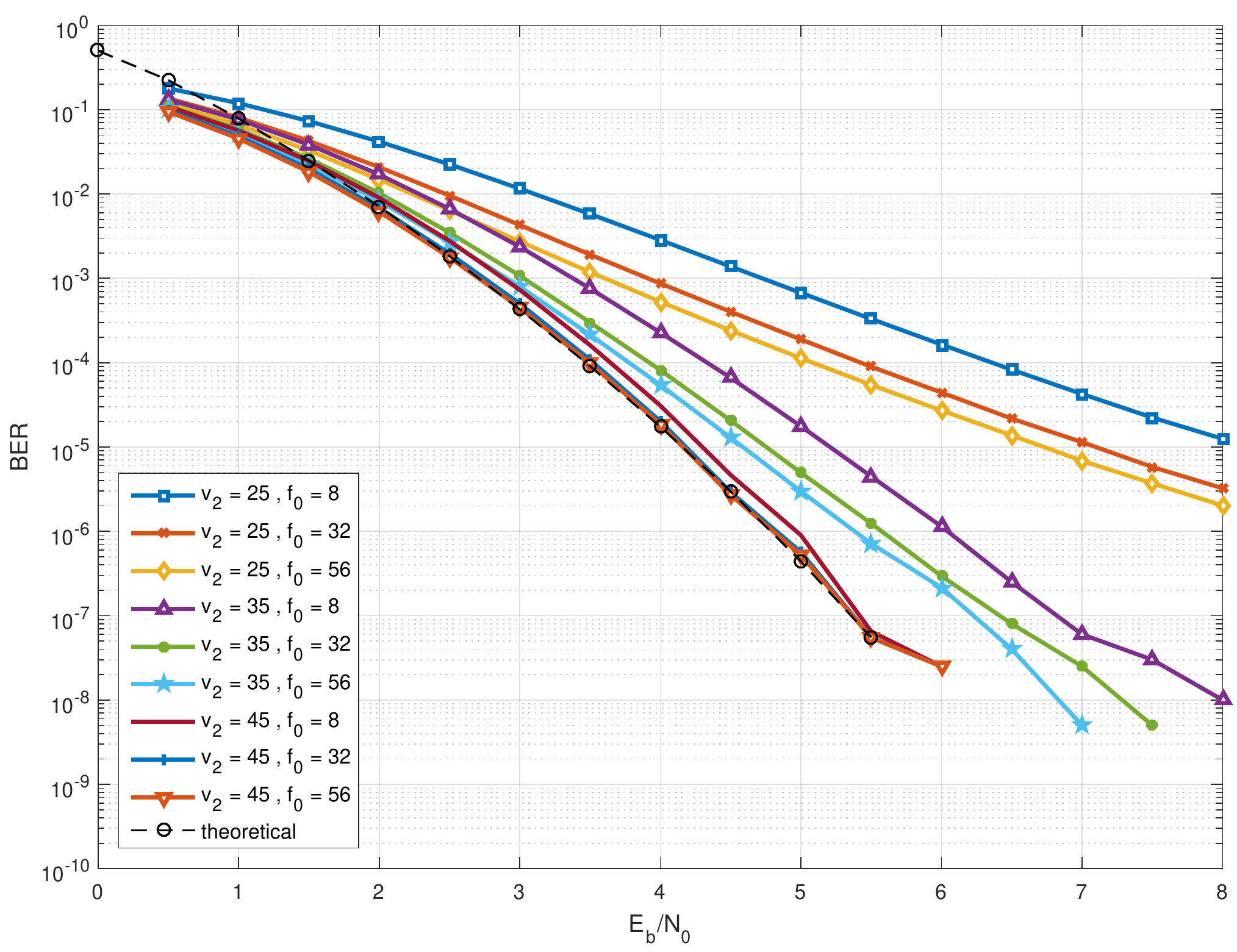} 
	\caption{The effect of $v_2$ and $f_0$ on BER in parallel traceback for $v_2=21$ and $f=300$}
	\label{fig:ber:pt}
\end{figure}

In parallel traceback, there is another parameter which is the non-overlapping size of subframes. Fig.~\ref{fig:ber:pt} shows the effect of both $v_2$ and $f_0$ on BER in parallel traceback algorithm. It can be seen that for $v_2=45$ and $f_0=32$, the decoder is reliable. It can also be observed that $v_2$ is still more important than the other parameters.

\begin{table}[t]
	\begin{center}
		\caption{The effect of $v_2$ and $f_0$ on BER in parallel taceback algorithm}
		\label{table:ber:pt}
		\begin{tabular}{|l||*{7}{l|}}
			\hline
			\backslashbox{$v_2$}{$f_0$} & \textbf{8} & \textbf{16} & \textbf{24} & \textbf{32} & \textbf{40} & \textbf{48} & \textbf{56} \\
			\hline
			\hline
			\textbf{25} &	2.90&	2.41&	2.15&	1.94&	1.77&	1.72&	1.54\\
			\hline
			\textbf{30} &	1.57&	1.28&	1.09&	0.97&	0.85&	0.81&	0.70\\
			\hline
			\textbf{35} &	0.87&	0.66&	0.53&	0.44&	0.39&	0.33&	0.29\\
			\hline
			\textbf{40} &	0.43&	0.31&	0.22&	0.18&	0.15&	0.12&	0.10\\
			\hline
			\textbf{45} &	0.18&	0.11&	0.08&	0.06&	0.05&	0.03&	0.03\\
			\hline
		\end{tabular}
	\end{center}
\end{table} 

In Table.~\ref{table:ber:pt} the metric discussed above is also used to show the effect of $v_2$ and $f_0$ on BER.

\begin{figure}[h]
	\centering
	\includegraphics[width=0.9\columnwidth]{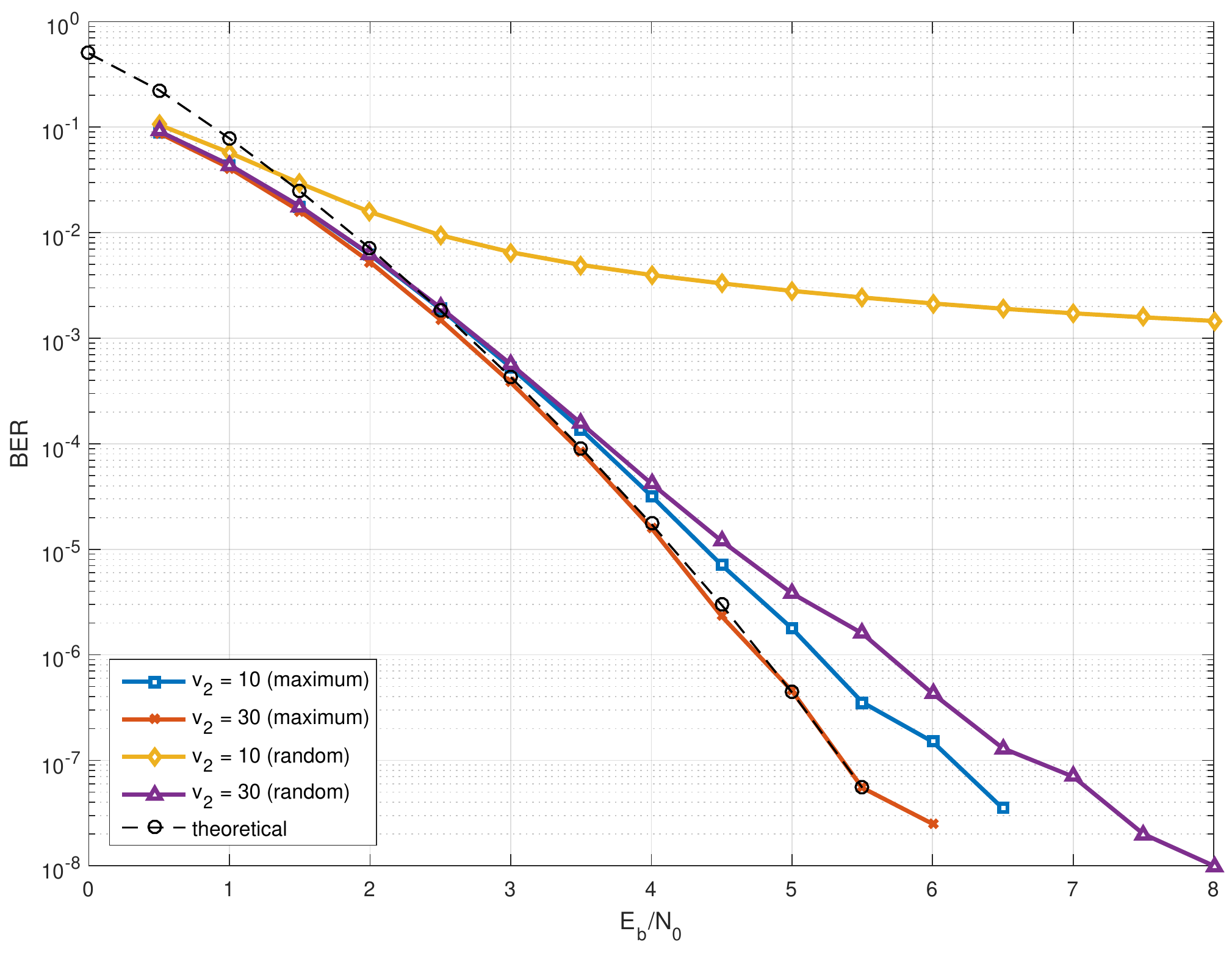} 
	\caption{The effect of traceback starting stated BER for $v_1=20$, $v_2=20$ and $f=256$}
	\label{fig:ber:sp}
\end{figure}

In parallel traceback algorithm, a challenge was explained regarding final path metrics of each subframe. Two discussed solutions were starting traceback from a random state and storing states with maximum path metric. Fig.~\ref{fig:ber:sp} shows that the first solution adversely affect BER and the cost of memory for storing the states pays off.

\begin{table}[t]
	\begin{center}
		\caption{Decoder throughput (Gb/s)}
		\label{table:thr}
		\begin{tabular}{|l||*{5}{l|}}
			\hline
			\backslashbox{$v_2$}{$f$} & \textbf{32} & \textbf{64} & \textbf{128} & \textbf{256} & \textbf{512} \\
			\hline
			\hline
			\textbf{10} &	4.28	&	5.11	&	6.64	&	6.15	&	4.97	\\
			\hline
			\textbf{20} &	3.79	&	4.79	&	6.36	&	6.05	&	4.86	\\
			\hline
			\textbf{30} &	3.10	&	4.23	&	5.74	&	5.77	&	4.80	\\
			\hline
			\textbf{40} &	2.82	&	3.93	&	5.50	&	5.62	&	4.77	\\
			\hline
		\end{tabular}
	\end{center}
\end{table}

\begin{table}[t]
	\begin{center}
		\caption{Decoder throughput (Gb/s) in parallel taceback.}
		\label{table:thr:pt}
		\begin{tabular}{|l||*{7}{l|}}
			\hline
			\backslashbox{$v_2$}{$f_0$} & \textbf{8} & \textbf{16} & \textbf{24} & \textbf{32} & \textbf{40} & \textbf{48} & \textbf{56} \\
			\hline
			\hline
			\textbf{25} &	12.1	&	11.7	&	13.7	&	11.9	&	13.5	&	12.4	&	13.0	\\
			\hline
			\textbf{30} &	10.2	&	10.0	&	12.1	&	10.3	&	11.9	&	10.9	&	11.5	\\
			\hline
			\textbf{35} &	8.47	&	8.47	&	10.6	&	8.79	&	10.3	&	9.45	&	9.95	\\
			\hline
			\textbf{40} &	6.74	&	7.11	&	9.15	&	7.37	&	8.82	&	8.00	&	8.48	\\
			\hline
			\textbf{45} &	4.95	&	5.28	&	7.58	&	5.84	&	7.23	&	6.39	&	6.83	\\
			\hline
		\end{tabular}
	\end{center}
\end{table} 

\subsection{Throughput}

Table.~\ref{table:thr} shows the throughput of the regular decoder over the same range of $f$ and $v_2$ as Table.~\ref{table:ber}. Likewise, Table.~\ref{table:thr:pt} that shows the throughput in parallel traceback algorithm is associated with Table.~\ref{table:ber:pt}. In order to demonstrate the improvement made by parallel traceback algorithm, cells of throughput should be compared. However, since parallel traceback increase BER, the cells that their associated cells in BER tables are close should be picked.

%



%




\ifCLASSOPTIONcaptionsoff
  \newpage
\fi


\begin{thebibliography}{10}
	\providecommand{\url}[1]{#1}
	\csname url@samestyle\endcsname
	\providecommand{\newblock}{\relax}
	\providecommand{\bibinfo}[2]{#2}
	\providecommand{\BIBentrySTDinterwordspacing}{\spaceskip=0pt\relax}
	\providecommand{\BIBentryALTinterwordstretchfactor}{4}
	\providecommand{\BIBentryALTinterwordspacing}{\spaceskip=\fontdimen2\font plus
		\BIBentryALTinterwordstretchfactor\fontdimen3\font minus
		\fontdimen4\font\relax}
	\providecommand{\BIBforeignlanguage}[2]{{%
			\expandafter\ifx\csname l@#1\endcsname\relax
			\typeout{** WARNING: IEEEtran.bst: No hyphenation pattern has been}%
			\typeout{** loaded for the language `#1'. Using the pattern for}%
			\typeout{** the default language instead.}%
			\else
			\language=\csname l@#1\endcsname
			\fi
			#2}}
	\providecommand{\BIBdecl}{\relax}
	\BIBdecl
	
	\bibitem{Viterbi}
	A.~{Viterbi}, ``Error bounds for convolutional codes and an asymptotically
	optimum decoding algorithm,'' \emph{IEEE Transactions on Information Theory},
	vol.~13, no.~2, pp. 260--269, April 1967.
	
	\bibitem{zhang2009}
	D.~Zhang, R.~Zhao, L.~Han, T.~Wang, and J.~Qu, ``An implementation of viterbi
	algorithm on gpu,'' in \emph{2009 First International Conference on
		Information Science and Engineering}.\hskip 1em plus 0.5em minus 0.4em\relax
	IEEE, 2009, pp. 121--124.
	
	\bibitem{kim2010ieeecomm}
	J.~Kim, S.~Hyeon, and S.~Choi, ``Implementation of an sdr system using graphics
	processing unit,'' \emph{IEEE communications magazine}, vol.~48, no.~3, pp.
	156--162, 2010.
	
	\bibitem{lin2011tiling}
	C.-S. Lin, W.-L. Liu, W.-T. Yeh, L.-W. Chang, W.-M.~W. Hwu, S.-J. Chen, and
	P.-A. Hsiung, ``A tiling-scheme viterbi decoder in software defined radio for
	gpus,'' in \emph{2011 7th International Conference on Wireless
		Communications, Networking and Mobile Computing}.\hskip 1em plus 0.5em minus
	0.4em\relax IEEE, 2011, pp. 1--4.
	
	\bibitem{gautam2014opencl}
	H.~Gautam, P.~Srinivasa, and S.~Kannan, ``Accelerating convolution coding \&
	viterbi decodingon gpus using opencl,'' in \emph{International Conference on
		Recent Advances and Innovations in Engineering (ICRAIE-2014)}.\hskip 1em plus
	0.5em minus 0.4em\relax IEEE, 2014, pp. 1--9.
	
	\bibitem{lee2013}
	K.-H. Lee and S.~W. Heo, ``Gpu based software dvb-t receiver design,'' in
	\emph{2013 IEEE International Conference on Consumer Electronics
		(ICCE)}.\hskip 1em plus 0.5em minus 0.4em\relax IEEE, 2013, pp. 582--585.
	
	\bibitem{lee2014}
	------, ``Openmp and gpu based software dvb-t receiver design,'' in \emph{2014
		IEEE International Conference on Consumer Electronics (ICCE)}.\hskip 1em plus
	0.5em minus 0.4em\relax IEEE, 2014, pp. 458--459.
	
	\bibitem{li2013}
	R.~Li, Y.~Dou, Y.~Li, and S.~Wang, ``A fully parallel truncated viterbi decoder
	for software defined radio on gpus,'' in \emph{2013 IEEE wireless
		communications and networking conference (WCNC)}.\hskip 1em plus 0.5em minus
	0.4em\relax IEEE, 2013, pp. 4305--4310.
	
	\bibitem{li2014}
	R.~Li, Y.~Dou, and D.~Zou, ``Efficient parallel implementation of three-point
	viterbi decoding algorithm on cpu, gpu, and fpga,'' \emph{Concurrency and
		Computation: Practice and Experience}, vol.~26, no.~3, pp. 821--840, 2014.
	
	\bibitem{peng2016}
	H.~Peng, R.~Liu, Y.~Hou, and L.~Zhao, ``A gb/s parallel block-based viterbi
	decoder for convolutional codes on gpu,'' in \emph{2016 8th International
		Conference on Wireless Communications \& Signal Processing (WCSP)}.\hskip 1em
	plus 0.5em minus 0.4em\relax IEEE, 2016, pp. 1--6.
	
\end{thebibliography}
\end{document}